\documentclass[a4paper,12pt]{article}

\pdfoutput=1

\usepackage[margin=1.5in]{geometry}
\usepackage{amsmath,amssymb,amsthm}
\usepackage{bm}
\usepackage{graphicx}
\usepackage[usenames, dvipsnames]{color}
\usepackage{natbib}
\usepackage{subcaption}
\usepackage{kpfonts}
\usepackage{hyperref}
\hypersetup{colorlinks=true,allcolors=blue}
\usepackage{authblk}

\setcitestyle{round}

\title{A comparison of methods for modeling marginal non-zero daily rainfall across the Australian continent}

\author[1]{Michael Bertolacci\thanks{Corresponding author, michael.bertolacci@research.uwa.edu.au}\textsuperscript{,}}
\author[1]{Edward Cripps}
\author[2]{Ori Rosen}
\author[3]{Sally Cripps}

\affil[1]{School of Mathematics and Statistics, University of Western Australia}
\affil[2]{Department of Mathematical Sciences, University of Texas at El Paso}
\affil[3]{Centre for Translational Data Science, University of Sydney}

\date{}

\newcommand*{\thetavec}{\bm{\theta}}
\newcommand*{\Ga}{\mathrm{Ga}}
\newcommand*{\nsites}{16,968}
\newcommand*{\nsitesnc}{16968}

\newcommand*{\pU}{{\color{red} U}}
\newcommand*{\pO}{{\color{red} O}}
\newcommand*{\pN}{{\color{ForestGreen} N}}

\begin{document}

\maketitle

\begin{abstract}
  \citet{naveau2016} have recently developed a class of methods, based on extreme-value theory (EVT), for capturing low, moderate, and heavy rainfall simultaneously, without the need to choose a threshold typical to EVT methods. We analyse the performance of Naveau et al.'s methods, along with mixtures of gamma distributions, by fitting them to marginal non-zero rainfall from \nsites{} sites spanning the Australian continent and which represent a wide variety of rainfall patterns. Performance is assessed by the distribution across sites of the log ratios of each method's estimated quantiles and the empirical quantiles. We do so for quantiles corresponding to low, moderate, and heavy rainfall. Under this metric, mixtures of three and four gamma distributions outperform Naveau et al's methods for small and moderate rainfall, and provide equivalent fits for heavy rainfall.
\end{abstract}

\section{Introduction}

This article compares seven methods for modeling marginal non-zero daily rainfall measurements.  Four of these methods are described in \citet{naveau2016}.  The remaining three methods comprise mixtures of two, three and four gamma densities.   The data are from  \nsites{} sites on the Australian continent and are reported in \citet{bertolacci2018}.

Performance for each method is assessed by its ability to estimate empirical quantiles. That is, for a given quantile, we examine the distribution across sites of the log ratio of the model's estimated quantile and the empirical quantile. This is done for the 0.01 and 0.1 quantiles (for low rainfall), the 0.25, 0.5 and 0.75 quantiles (for moderate rainfall), and the 0.9 and 0.99 quantiles (for heavy rainfall).

Under this metric, for low and moderate rainfall, the mixtures of three and four gamma densities outperform the four methods from \citet{naveau2016}.  For heavy rainfall, the 0.9 quantile is again best estimated by the three and four gamma mixtures.  \citet{naveau2016}'s method of probability weighted moments with censoring performs best for the 0.99 quantile, although  the improvement over the mixture of four gamma densities is marginal.

\section{Data}
\label{sec:data}

The dataset of \citet{bertolacci2018} comprises daily rainfall measurements collected by the Australian Bureau of Meteorology (BOM) at \nsites{} observation sites across the Australian continent. The BOM describes rainfall measurements at its observational sites as:\footnote{\url{http://www.bom.gov.au/climate/cdo/about/about-rain-data.shtml}}
\begin{quote}
  Rainfall includes all forms of water particles, whether liquid (for example, rain or drizzle) or solid (hail or snow), that fall from clouds and reaches the ground. The rain gauge is the standard instrument for recording rainfall, which is measured in millimetres. Rainfall is generally observed daily at 9 am local time---this is a measure of the total rainfall that has been received over the previous 24 hours.
\end{quote}

\begin{figure}[ht]
  \centering
  \begin{subfigure}[t]{0.49\textwidth}
    \centering
    \includegraphics[width=\linewidth]{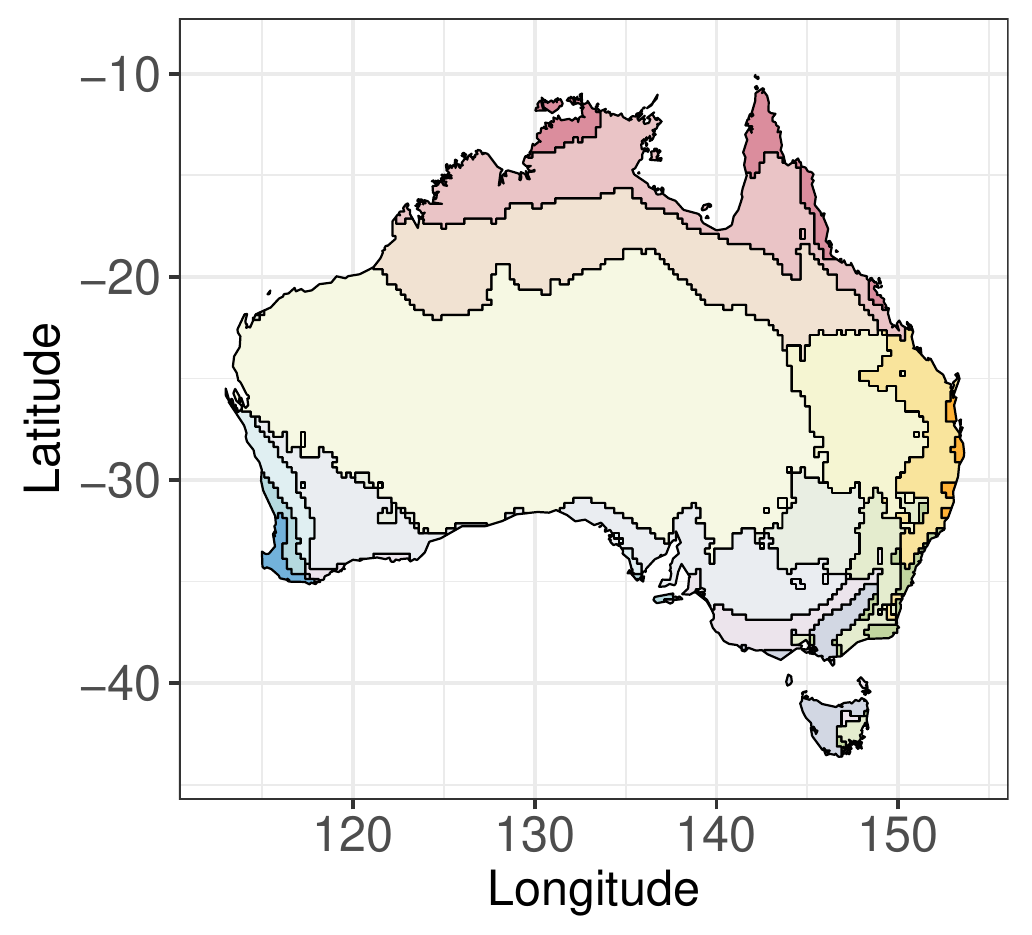}
    \caption{}
    \label{fig:aust_climate_classifications}
  \end{subfigure}
  \begin{subfigure}[t]{0.49\textwidth}
    \centering
    \includegraphics[width=\linewidth]{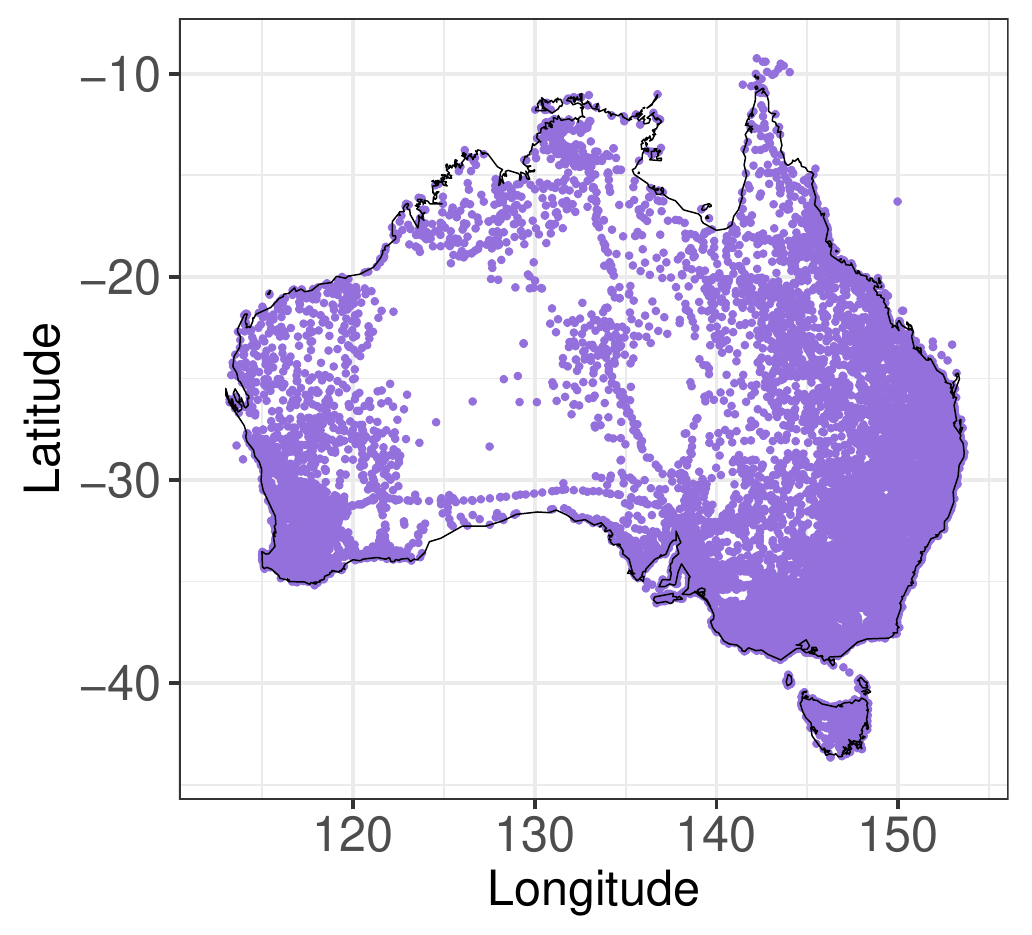}
    \caption{}
    \label{fig:aust_sites}
  \end{subfigure}

  \includegraphics[width=\linewidth]{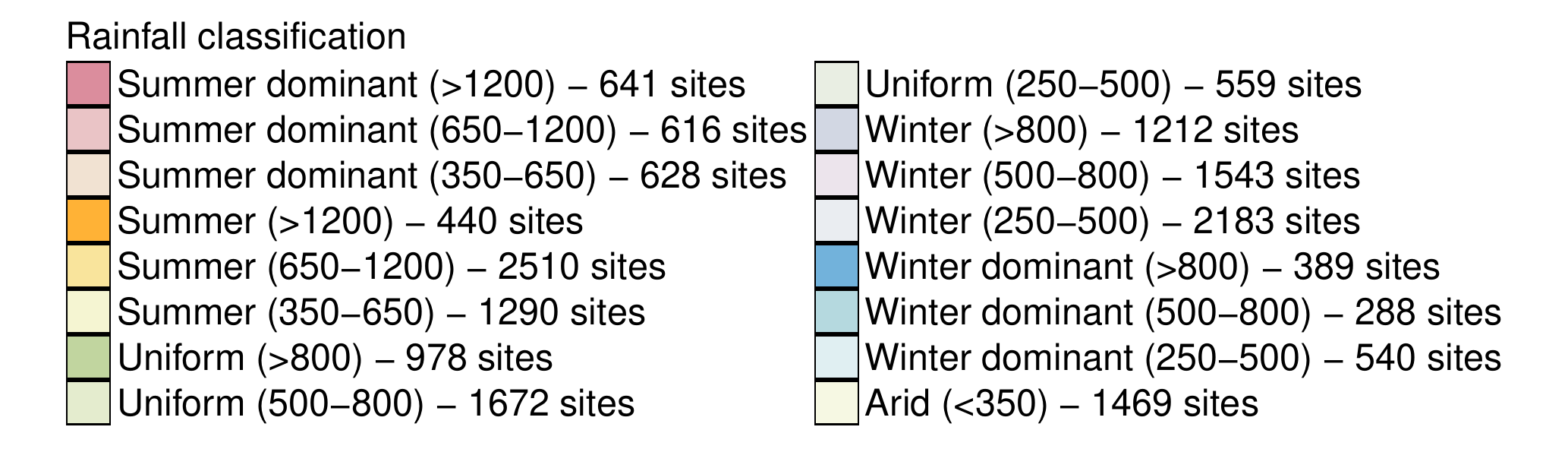}

  \caption{Australian Bureau of Meteorology (BOM) rainfall categories with median annual rainfall levels in millimeters (a) and  the locations of the \nsites{} rainfall observation sites analyzed in this article (b).}
  \label{fig:aust_panel}
\end{figure}

The dataset ranges across a wide geographical area comprising the entire Australian continent, with a correspondingly wide range of rainfall patterns. Climate classifications by the BOM for the continent are depicted in Figure~\ref{fig:aust_climate_classifications}. Regions are classified based on the season of maximum rainfall (summer, winter, uniform, or arid), how marked is the seasonality (summer/winter dominant or not dominant), and by median annual rainfall. The locations of the \nsites{} sites are pictured in Figure~\ref{fig:aust_sites}, all of which recorded at least 100 days of non-zero rainfall.

\begin{figure}
  \centering
  \begin{subfigure}[t]{\textwidth}
    \centering
    \includegraphics[width=\linewidth]{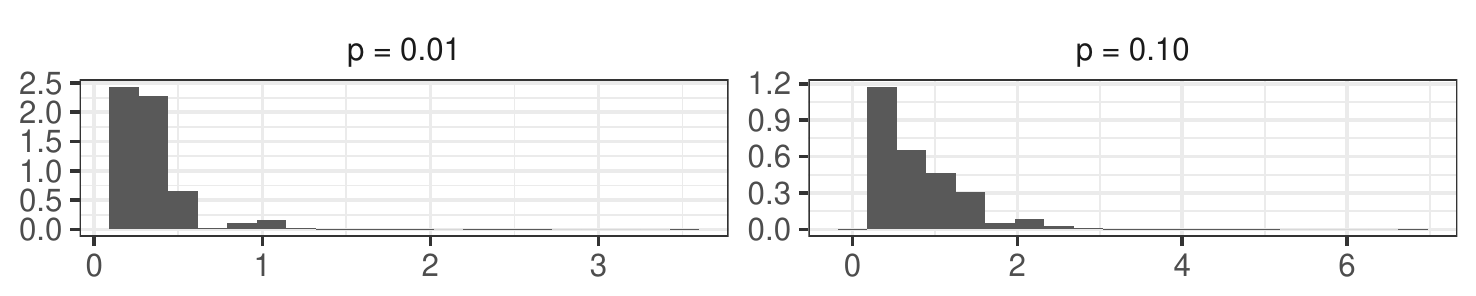}
    \caption{Low rainfall}
    \label{fig:empirical_quantile_histogram_low}
  \end{subfigure}
  \begin{subfigure}[t]{\textwidth}
    \centering
    \includegraphics[width=\linewidth]{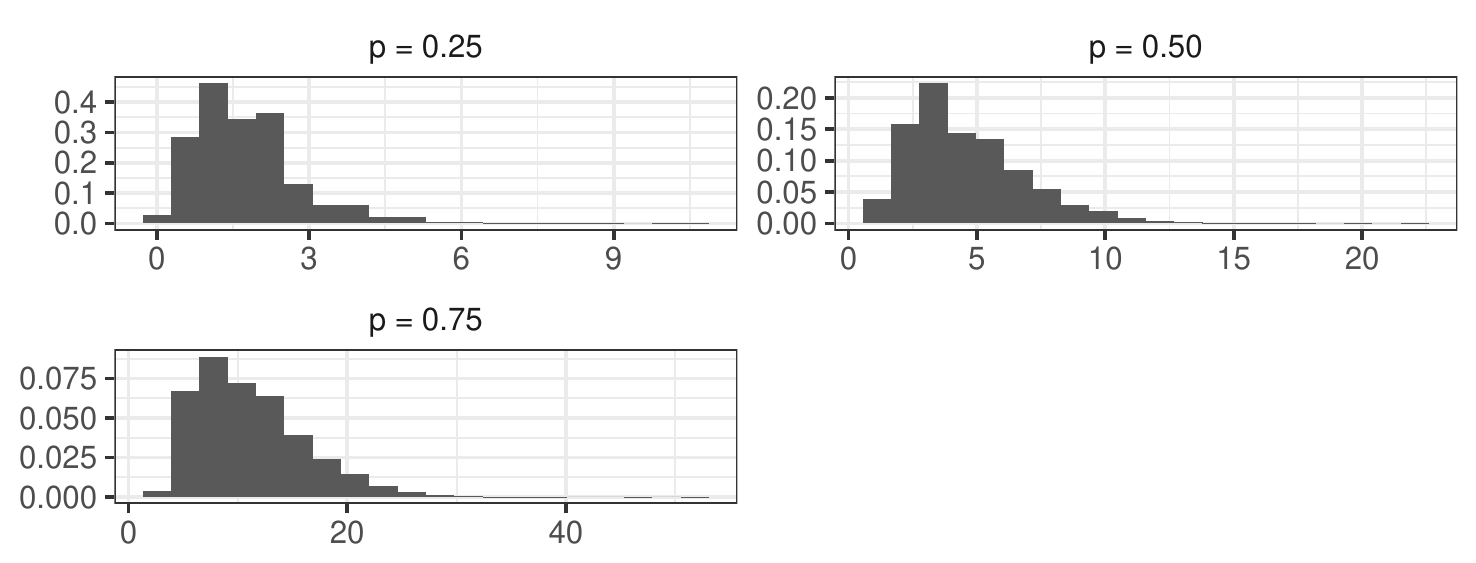}
    \caption{Moderate rainfall}
    \label{fig:empirical_quantile_histogram_moderate}
  \end{subfigure}
  \begin{subfigure}[t]{\textwidth}
    \centering
    \includegraphics[width=\linewidth]{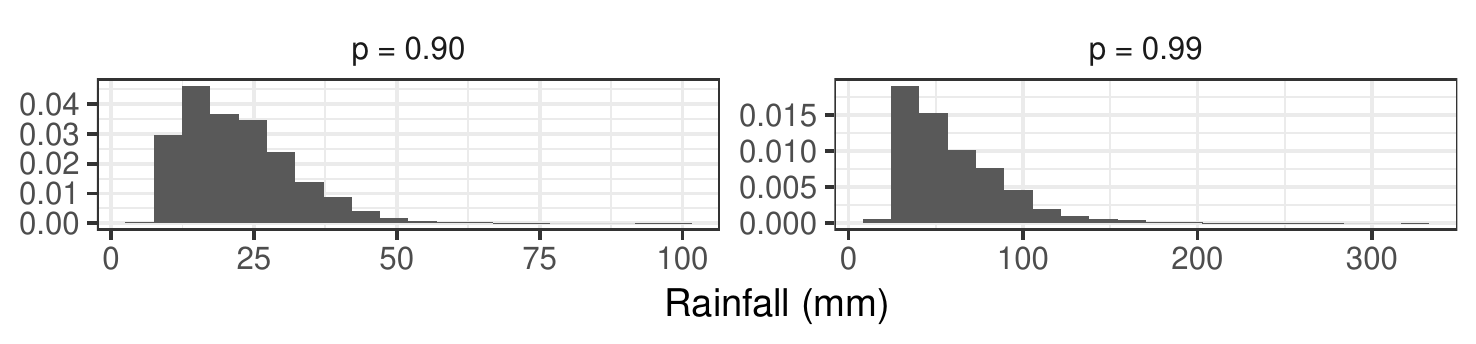}
    \caption{Heavy rainfall}
    \label{fig:empirical_quantile_histogram_heavy}
  \end{subfigure}

  \caption{Density histograms of empirical quantiles, $\hat{q}_\text{e}^p(s)$, across all of the \nsites{} sites for low (top), moderate (middle), and heavy (bottom) rainfall quantiles.}
  \label{fig:empirical_quantile_histogram}
\end{figure}

Define $\hat{q}_\text{e}^p(s)$ as the empirical $p$th quantile for marginal non-zero rainfall at site $s$. Density histograms of $\hat{q}_\text{e}^p(s)$ for low rainfall at $p = 0.01, 0.1$ are shown in Figure~\ref{fig:empirical_quantile_histogram_low}, moderate rainfall at $p=0.25, 0.50, 0.75$ in Figure~\ref{fig:empirical_quantile_histogram_moderate}, and heavy rainfall at $p=0.90, 0.99$ in Figure~\ref{fig:empirical_quantile_histogram_heavy}. These histograms show that marginal rainfall quantiles vary widely across Australia.

\section{Methods investigated}
\label{sec:methods_investigated}

\subsection{Generalised Pareto distribution of \texorpdfstring{\citet{naveau2016}}{Naveau et al. (2016)}}

\citet{naveau2016} extend Extreme-Value Theory (EVT) to derive a simple parametric model that jointly accommodates low, moderate, and heavy rainfall. This model can be specified via the CDF

\begin{equation}
  F(y) = G\left( H\left( y; \sigma, \xi \right); \thetavec \right),
\end{equation}
where $H(y; \sigma, \xi)$ is the CDF of the Generalized Pareto distribution with parameters $\sigma$ and $\xi$, and $G(x; \thetavec)$ is a CDF on $[0, 1]$ parametrized by $\thetavec$.  \citet{naveau2016}  present four candidate forms of $G$ and indicate that $G(x; \kappa) = x^\kappa$, $\kappa > 0$, gives the best overall performance. In the remainder of this article, we use $G(x; \kappa) = x^\kappa$ and the following four methods outlined in \citet{naveau2016} to estimate the parameters of $F(y)$:

\begin{itemize}
  \item
    \textsc{Naveau-MLE}, which fits the model using maximum likelihood estimation (MLE);

  \item
    \textsc{Naveau-PWM}, which fits the model using probability weighted moments (PWM);

  \item
    \textsc{Naveau-MLE-c}, which fits the model using MLE, where values below a threshold $y < y_L$ are censored;

  \item
    \textsc{Naveau-PWM-c}, which fits the model using PWM, where values $y < y_L$ are censored.
\end{itemize}

The details for each of these methods are given by \citet{naveau2016}, who find that censoring provides better model fits to heavy rainfall quantiles. They attribute this to the amelioration of the impact of the discretization of values introduced by limited measurement precision. For this work, we set the threshold $y_L = 1$mm.   Estimation is performed using the \texttt{egp2} function in the R package \texttt{mev} \citep{rcore2016,mev2017}, which was written by the first two authors of \citet{naveau2016}.

\subsection{Mixtures of \texorpdfstring{$K$}{K} Gamma distributions}

This article also evaluates fits of mixtures of $K \geq 2$ gamma distributions, which we denote \textsc{Gamma-Mixture-K}. For this model,

\begin{equation}
  y_i \sim \sum_{k = 1}^K \pi_k \Ga(a_k, b_k),
\end{equation}
where $0 \leq \pi_k \leq 1$, $\sum_{k = 1}^K \pi_k = 1$, and $\Ga(a, b)$ is a gamma distribution with density $f(y) \propto y^{a - 1} e^{-y / b}$. Following \citet{bertolacci2018}, we use the conjugate prior of \citet{damsleth1975} for $a_k$ and $b_k$, with $b_k \sim \mathrm{IG}(u, v)$ and $p(a_k \mid b_k) \propto \frac{\rho^{a_k - 1}}{b_k^{a_k q} \Gamma(a_k)^{r}}$, where $u = 1.1, v = 2$, and $\rho = q = r = 1$.

We estimate the parameters of the \textsc{Gamma-Mixture-K} model using the maximum a posteriori (MAP) method, implemented in the probabilistic programming language Stan and using the R package RStan \citep{carpenter2017,rstan2018}.

\section{Results}

Denote by $\hat{q}_m^p(s)$ the estimated $p$th quantile for method $m$ at site $s$.  As a measure of correspondence between $\hat{q}_m^p(s)$ and $\hat{q}_e^p(s)$ we write
\begin{equation}
  D^p_m(s) = \log \frac{\hat{q}_m^p(s)}{\hat{q}_\text{e}^p(s)}.
  \label{eqn:metric}
\end{equation}
When $D^p_m(s)$ is 0, $\hat{q}_m^p(s)$ and $\hat{q}_e^p(s)$ are equal, while if the method over/under-estimates the empirical quantile, $D^p_m(s)$ is positive/negative, with larger absolute values indicating poorer performance.

\begin{figure}
  \centering
  \includegraphics[width=\linewidth]{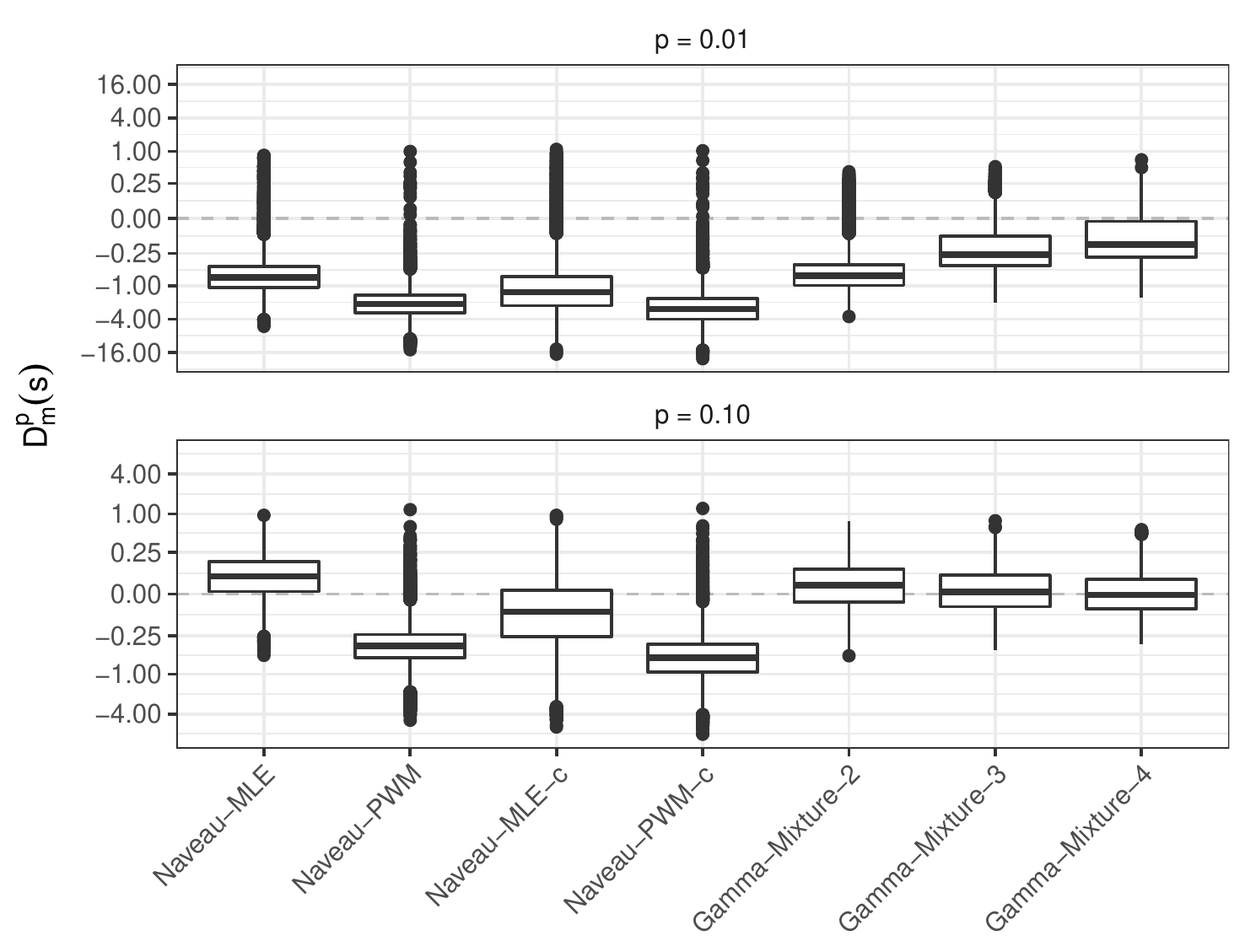}

  \caption{Boxplots of $D^p_m(s)$ for the low rainfall quantiles $p = 0.01$ (top) and $p = 0.10$ (bottom), across all $s = 1, 2, \ldots, \nsitesnc{}$ sites, for each method $m$. The value 0, corresponding to perfect fit, is indicated by a dashed line. The scale is transformed by $\mathrm{asinh}(8x)$.}
  \label{fig:quantile_boxplots_low}
\end{figure}

\begin{figure}
  \centering
  \includegraphics[width=\linewidth]{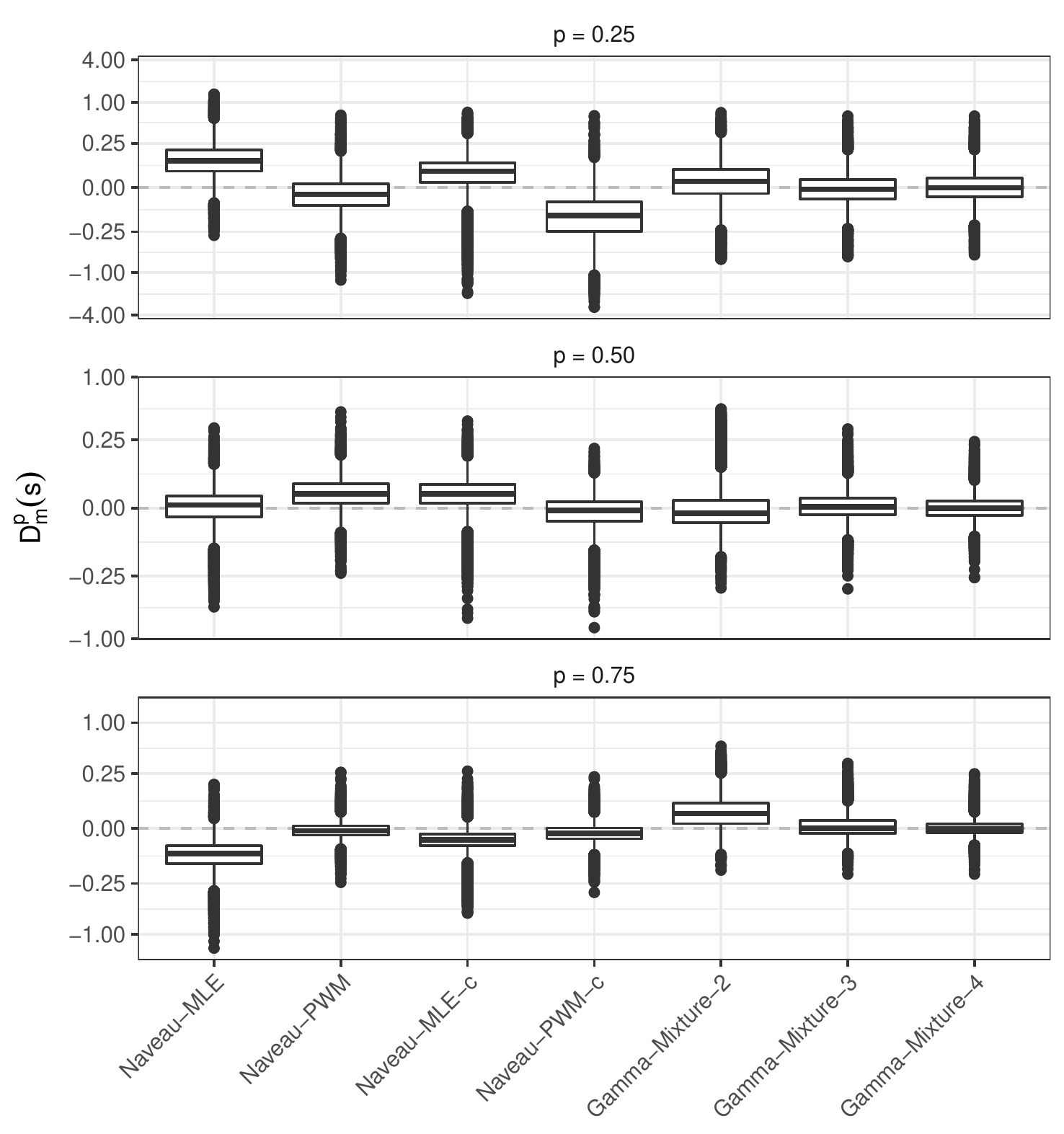}

  \caption{Boxplots of $D^p_m(s)$ for the moderate rainfall quantiles $p = 0.25$ (top), $p = 0.50$ (middle), and $p = 0.75$ (bottom), across all $s = 1, 2, \ldots, \nsitesnc{}$ sites, for each method $m$. The value 0, corresponding to perfect fit, is indicated by a dashed line. The scale is transformed by $\mathrm{asinh}(8x)$.}
  \label{fig:quantile_boxplots_moderate}
\end{figure}

\begin{figure}
  \centering
  \includegraphics[width=\linewidth]{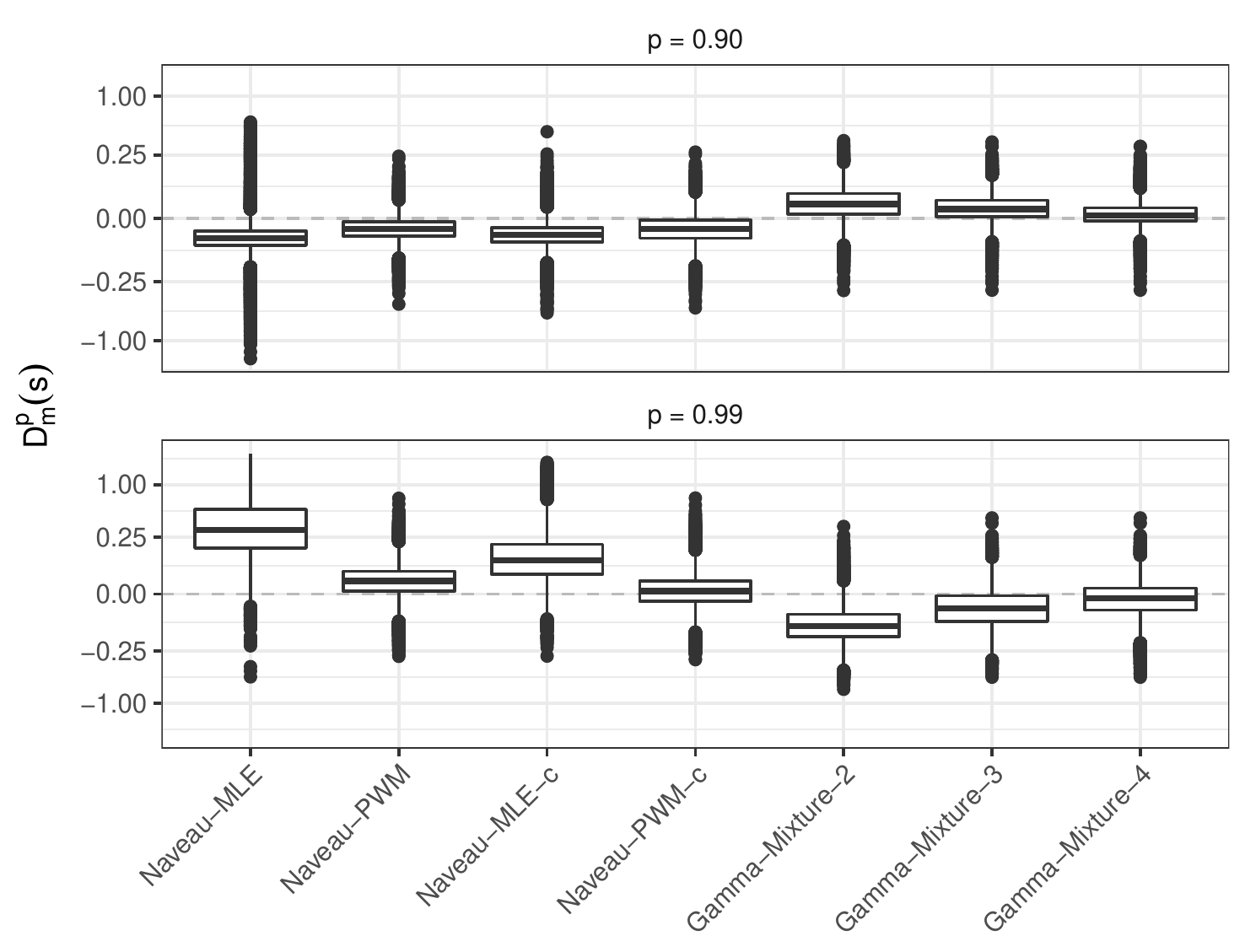}

  \caption{Boxplots of $D^p_m(s)$ for the heavy rainfall quantiles $p = 0.90$ (top) and $p = 0.99$ (bottom), across all $s = 1, 2, \ldots, \nsitesnc{}$ sites, for each method $m$. The value 0, corresponding to perfect fit, is indicated by a dashed line. The scale is transformed by $\mathrm{asinh}(8x)$.}
  \label{fig:quantile_boxplots_heavy}
\end{figure}

We apply seven methods (described in Section~\ref{sec:methods_investigated}), \textsc{Naveau-MLE}, \textsc{Naveau-PWM}, \textsc{Naveau-MLE-c}, \textsc{Naveau-PWM-c}, \textsc{Gamma-Mixture-2}, \textsc{Gamma-Mixture-3} and \textsc{Gamma-Mixture-4}, to each of the \nsites{} sites from Section~\ref{sec:data}, for a total of 118,776 separate model fits. For each method $m$ and quantile $p$, we assess the results via the distribution of $D^p_m(s)$ across all sites $s$. Figures~\ref{fig:quantile_boxplots_low}, \ref{fig:quantile_boxplots_moderate}, and \ref{fig:quantile_boxplots_heavy} show boxplots of $D^p_m(s)$ for the low, moderate, and heavy rainfall quantiles, respectively, with one boxplot for each combination of method $m$ and $p$.   Table~\ref{tab:median_summary} gives the median of $D^p_m(s)$ for each method/quantile. Finally, the interquartile range (IQR) of $D^p_m(s)$ is used to classify each method as underestimating ($\mathrm{IQR} < 0$), overestimating ($\mathrm{IQR} > 0$), or nominal ($\mathrm{IQR}$ contains $0$).  Table~\ref{tab:over_under_summary} gives the classification as underestimating (shown with a \pU{}), overestimating (\pO{}), or nominal (\pN{}), for each method/quantile.

Figure~\ref{fig:quantile_boxplots_low} shows that for both $p = 0.01$ and $p = 0.10$, \textsc{Gamma-Mixture-2}, \textsc{-3}, and \textsc{-4} perform best. When $p = 0.01$ all methods underestimate. Median values of $D^p_m(s)$ (Table~\ref{tab:median_summary}) show that \textsc{Naveau-PWM-c} underestimates the most, while \textsc{Gamma-Mixture-4} underestimates the least. The other low rainfall quantile, $p = 0.10$ (Figure~\ref{fig:quantile_boxplots_low}, bottom), is also the most underestimated by \textsc{Naveau-PWM-c}, but is estimated nominally and best by \textsc{Gamma-Mixture-4}. The median values of $D^p_m(s)$ for each method when $p = 0.10$ have lower absolute value than those for $p = 0.01$, indicating that this quantile is better captured in general. For both $p = 0.01$ and $p = 0.10$, mixtures of gammas outperform the methods of Naveau et al.

Figure~\ref{fig:quantile_boxplots_moderate} shows that when $p = 0.25$, \textsc{Gamma-Mixture-4} performs best and is nominal, while \textsc{Naveau-PWM-c} and \textsc{-MLE}, which under- and overestimate, respectively, perform worst. When $p = 0.50$, \textsc{Gamma-Mixture-4} performs best (Figure~\ref{fig:quantile_boxplots_moderate}, middle) and all methods are nominal, except for \textsc{Naveau-PWM} and \textsc{-MLE-c} which both overestimate. When $p = 0.75$ (Figure~\ref{fig:quantile_boxplots_moderate}, bottom), \textsc{Naveau-MLE} and \textsc{Gamma-Mixture-2}, which under- and overestimate, respectively, perform worst, while it is best and nominally estimated by \textsc{Gamma-Mixture-3} and \textsc{-4}.

The heavy rainfall quantile, $p = 0.90$, (Figure~\ref{fig:quantile_boxplots_heavy}, top), is estimated nominally by \textsc{Gamma-Mixture-4}, underestimated by all \textsc{Naveau-} methods, and overestimated by \textsc{Gamma-Mixture-2} and \textsc{-3}. Median values of $D^p_m(s)$ show that \textsc{Naveau-MLE} and \textsc{-MLE-c} give the worst fit, while \textsc{Gamma-Mixture-4} gives the best fit. Finally, the heaviest rainfall quantile, $p = 0.99$ (Figure~\ref{fig:quantile_boxplots_heavy}, bottom), is estimated nominally by \textsc{Naveau-PWM-c} and \textsc{Gamma-Mixture-4}, overestimated by the remaining \textsc{Naveau-} methods, and underestimated by \textsc{Gamma-Mixture-2} and \textsc{-3}. The best fit is given by \textsc{Naveau-PWM-c}, which shows slight improvement over \textsc{Gamma-Mixture-4}, while \textsc{Naveau-MLE} performs worst.

Table~\ref{tab:median_summary} shows that the worst performance for all methods is when $p = 0.01$, while the moderate rainfall quantiles $p = 0.25, 0.50$ and $0.75$ are generally well estimated.  Among the \textsc{Naveau-} fits, the PWM methods outperform the MLE methods for the heavy rainfall quantiles $p = 0.90$ and $0.99$, replicating the findings of \citet{naveau2016} that PWM is more robust for estimating extreme heavy rainfall. However, the situation is reversed for the low rainfall quantiles $p = 0.01$ and $0.10$, with the MLE methods having lower median values of $D^p_m(s)$. The same pattern exists for the censoring methods, which perform better for heavy rainfall, and worse for low rainfall.

Overall, the method that performs best is \textsc{Gamma-Mixture-4}, followed by \textsc{Gamma-Mixture-3}.  Table~\ref{tab:over_under_summary} reports nominal performance for \textsc{Gamma-Mixture-4} at all quantiles except the $0.01$ quantile, although it outperforms the other methods for this quantile (see Table~\ref{tab:median_summary}). Table~\ref{tab:median_summary} reports the lowest absolute values of the median $\hat{D}^p_m(s)$ for \textsc{Gamma-Mixture-4} for all quantiles except $p=0.99$, where \textsc{Naveau-PWM-c} marginally improves upon \textsc{Gamma-Mixture-4}. The second best, \textsc{Gamma-Mixture-3}, has the second lowest median values for all quantiles except $p = 0.99$. Among the \textsc{Naveau-} methods, \textsc{Naveau-PWM-c} is most frequently nominal, with the caveat that it is also the worst performing method for the $0.01, 0.10$ and $0.25$ quantiles.

\begin{table}[ht]
\centering
{\small
\begin{tabular}{l|rr|rrr|rr}
                          & \multicolumn{7}{c}{Quantile ($p$)} \\
                          & \multicolumn{2}{c|}{Low} & \multicolumn{3}{c|}{Moderate} & \multicolumn{2}{c}{Heavy} \\
Method ($m$)              & $0.01$ & $0.10$ & $0.25$ & $0.50$ & $0.75$ & $0.90$ & $0.99$ \\ \hline
\textsc{Naveau-MLE} & -709.2 & 82.4 & 123.2 & 7.8 & -87.7 & -58.3 & 304.4 \\
\textsc{Naveau-PWM} & -2139.2 & -367.4 & -27.0 & 38.3 & -7.1 & -30.4 & 41.9 \\
\textsc{Naveau-MLE}-c & -1301.8 & -82.7 & 69.3 & 38.6 & -37.4 & -48.4 & 121.2 \\
\textsc{Naveau-PWM}-c & -2646.4 & -565.2 & -130.1 & -6.7 & -15.6 & -30.4 & \textbf{8.7} \\
\textsc{Gamma-Mixture-2} & -661.8 & 39.3 & 26.5 & -13.6 & 50.3 & 42.0 & -113.1 \\
\textsc{Gamma-Mixture-3} & -263.4 & 8.9 & -7.2 & 3.6 & 1.0 & 26.8 & -46.3 \\
\textsc{Gamma-Mixture-4} & \textbf{-162.3} & \textbf{-4.3} & \textbf{-0.7} & \textbf{-0.7} & \textbf{-0.2} & \textbf{8.8} & -14.1
\end{tabular}
}

\caption{Median value of $D^p_m(s)$ of each method $m$ and quantile $p$. All values are $\times 10^{-3}$. The lowest magnitude value in each column is in \textbf{bold}.}
\label{tab:median_summary}
\end{table}

\begin{table}
\centering
{\small
\begin{tabular}{l|rr|rrr|rr}
                          & \multicolumn{7}{c}{Quantile ($p$)} \\
                          & \multicolumn{2}{c|}{Low} & \multicolumn{3}{c|}{Moderate} & \multicolumn{2}{c}{Heavy} \\
Method ($m$)              & $0.01$ & $0.10$ & $0.25$ & $0.50$ & $0.75$ & $0.90$ & $0.99$ \\
\hline
\textsc{Naveau-MLE}       & \pU{}  & \pO{}   & \pO{}   & \pN{}     & \pU{}  & \pU{}  & \pO{}   \\
\textsc{Naveau-PWM}       & \pU{}  & \pU{}  & \pN{}     & \pO{}   & \pN{}     & \pU{}  & \pO{}   \\
\textsc{Naveau-MLE-c}     & \pU{}  & \pN{}     & \pO{}   & \pO{}   & \pU{}  & \pU{}  & \pO{}   \\
\textsc{Naveau-PWM-c}     & \pU{}  & \pU{}  & \pU{}  & \pN{}     & \pN{}     & \pU{}  & \pN{}     \\
\textsc{Gamma-Mixture-2}  & \pU{}  & \pN{}     & \pN{}     & \pN{}     & \pO{}   & \pO{}   & \pU{}  \\
\textsc{Gamma-Mixture-3}  & \pU{}  & \pN{}     & \pN{}     & \pN{}     & \pN{}     & \pO{}   & \pU{}  \\
\textsc{Gamma-Mixture-4}  & \pU{}  & \pN{}     & \pN{}     & \pN{}     & \pN{}     & \pN{}     & \pN{}     \\
\end{tabular}
}

\caption{Summary of the performance for each method $m$ and quantile $p$, based on the distribution of $D^p_m(s)$. `\pU{}' stands for underestimated, defined as the interquartile range (IQR) of $D^p_m(s)$ less than $0$, `\pO{}' stands for overestimated, with the IQR above zero, and `\pN{}' stands for nominal, with the IQR containing zero.}
\label{tab:over_under_summary}
\end{table}

\section{Conclusion}

We have presented a comparison of methods described by \citet{naveau2016}, and mixtures of gamma distributions, for fitting marginal non-zero daily rainfall. The data were obtained from \nsites{} sites spanning the Australian continent, and representing a wide variety of marginal rainfall patterns. The performance of each method was assessed by its ability to estimate empirical quantiles, at the 0.01 and 0.1 quantiles for low rainfall, the 0.25, 0.5 and 0.75 quantiles for moderate rainfall, and the 0.9 and 0.99 quantiles for heavy rainfall. The method that performs best is the mixture of four gamma distributions, being slightly outperformed only on the $0.99$th quantile by one of Naveau et al.'s methods. In general, mixtures of three and four gamma distributions outperform other methods for small and moderate rainfall, and provide better or equivalent fits for heavy rainfall.

Using \citet{naveau2016}'s software, we replicate their finding that probability weighted moments estimation for their class of models outperforms maximum likelihood for the heavy rainfall quantiles, and augment this, finding that the situation is reversed for the low rainfall quantiles. The same pattern exists for censoring based methods, with censoring providing better/worse fits for heavy/low rainfall, respectively.

An interesting result, common to all methods tested, is the underestimation of the low rainfall $0.01$th quantile. This quantile performs the worst for all methods, which is surprising given that the rainfall literature has largely focussed on estimating heavy rainfall quantiles. The underestimation of this quantile may be an artefact of the discretisation of the data due to instrument precision, but this remains to be confirmed, and is a subject for future research. In any case, mixtures of gammas provide better fits than the methods of Naveau et al. for low rainfall for both the $0.01$th and the $0.10$th quantile, with the fits improving as more mixture components are added.

\bibliographystyle{plainnat}
\bibliography{marginal_rainfall_paper}{}

\end{document}